\documentclass[superscriptaddress,twocolumn,amsmath,amssymb,nofootinbib,longbibliography]{revtex4-1}

\newcommand{\tens}[1]{{\boldsymbol{#1}}}       
\newcommand{\ts}[1]{{\boldsymbol{#1}}}         

\newcommand{\grad}{{\tens{d}}}                 
\newcommand{\lied}{\pounds}                    
\newcommand{\cv}[1]{{\tens{\partial}}_{#1}}    
\newcommand{\pa}{\partial}                     

\newcommand{\kt}{K}                                            
\newcommand{\ktg}{k}                                           

\newcommand{\Ag}{A}                                            

\newcommand{\EMA}{{A}}                             
\newcommand{\EMF}{{F}}                             

\newcommand{\EMAbgr}{{\mathcal{A}}}                 
\newcommand{\EMFbgr}{{\mathcal{F}}}                 

\newcommand{\be}{\begin{equation}}             
\newcommand{\ee}{\end{equation}}               
\newcommand{\ba}{\begin{eqnarray}}             
\newcommand{\ea}{\end{eqnarray}}               
\newcommand{\n}[1]{\label{#1}}
\newcommand{\hh}{,\hspace{0.5cm}}
\newcommand{\hhh}{,\hspace{0.2cm}}

\begin{document}

\title{Separation of variables in Maxwell equations in Pleba\'nski--Demia\'nski spacetime}

\author{Valeri P. Frolov}
\email{vfrolov@phys.ualberta.ca}
\affiliation{Theoretical Physics Institute, University of Alberta, Edmonton,
Alberta, Canada T6G 2E1}
\affiliation{Yukawa Institute for the Theoretical Physics, Kyoto University, 606-8502, Kyoto, Japan}
\author{Pavel Krtou\v{s}}
\email{Pavel.Krtous@utf.mff.cuni.cz}
\affiliation{Institute of Theoretical Physics, Faculty of Mathematics and Physics, Charles University,\\
V~Hole\v{s}ovi\v{c}k\'ach~2, Prague, 18000, Czech Republic}
\author{David Kubiz\v{n}\'ak}
\email{dkubiznak@perimeterinstitute.ca}
\affiliation{Perimeter Institute, 31 Caroline Street North, Waterloo, ON, N2L 2Y5, Canada}

\date{February 25, 2018}         

\begin{abstract}
A new method for separating variables in Maxwell's equations in four- and higher-dimensional  Kerr--(A)dS spacetimes proposed recently by Lunin is generalized to any off-shell metric that admits a principal Killing--Yano tensor. The key observation is that  Lunin's ansatz for the vector potential can be formulated in a covariant form---in terms of the principal tensor. In particular, focusing on the four-dimensional case we demonstrate separability of Maxwell's equations in  the Kerr--NUT--(A)dS and the Pleba\'nski--Demia\'nski family of spacetimes. The new method of separation of variables is quite different from the standard approach based on the Newman--Penrose formalism.
\end{abstract}

\maketitle

\section{Introduction}

A study of electromagnetic fields in the vicinity of (rotating) black holes in four dimensions yields interesting astrophysical applications and has been investigated by many authors, see e.g. \cite{Wald:1974np,King:1975tt, bicak1977stationary, bivcak1976stationary, bivcak1980stationary, bivcak1985magnetic, Aliev:1989wx, Penna:2014aza}. See also \cite{Aliev:2004ec, Ortaggio:2004kr, Aliev:2006tt,Ortaggio:2006ng,  Aliev:2007qi, Krtous:2007,Frolov:2010cr} for the studies in higher dimensions.

In the test field approximation, that is for a small field amplitude and when the corresponding backreaction on the metric can be neglected, the electromagnetic field is described by a solution of linear Maxwell equations in a given metric.
The remarkable fact discovered by Teukolsky \cite{Teukolsky:1972,Teukolsky:1973} is that for
a vacuum rotating black hole, described by the Kerr geometry \cite{Kerr:1963},
the Maxwell equations can be decoupled and the resulting scalar (master) equations admit complete separation of variables. This result was later generalized by Torres del Castillo \cite{TorresdelCastillo:1988} to a general class of type D electrovacuum spacetimes described by the Pleba\'nski--Demia\'nski metric \cite{PlebanskiDemianski:1976}. Both these results were derived by employing the  Newman--Penrose formalism \cite{NewmanPenrose:1962,PenroseRindler:book,Chandrasekhar:1983}.

The separation of variables in higher-dimensional Maxwell equations remained an open problem for a long time. A partial success was achieved in \cite{Araneda:2017mdk} where it was demonstrated that such equations can be decoupled, using the higher-dimensional generalization of the Newman--Penrose formalism \cite{Ortaggio:2012jd}, provided that the background spacetime is Kundt, i.e., it admits a null geodesic congruence which is all: shear-free, twist-free, and expansion-free. Unfortunately, the higher-dimensional rotating black hole spacetimes do not belong to this class. However, as shown in \cite{Araneda:2017mdk} the results can be applied to some special limiting cases, including in particular the near horizon geometries of extremal black holes (see also \cite{Durkee:2010qu, Durkee:2010ea}).

A remarkable progress regarding the separability of the Maxwell equations in rotating black hole spacetimes has been recently achieved by Lunin \cite{Lunin:2017}. In his paper, Lunin proposed a new ansatz for the vector potential (rather than the field strength as customary in the Newman--Penrose formalism) and showed that it admits a separation of variables. The new method works equally in four and higher dimensions---the separability of Maxwell's equations in the four-dimensional Kerr-(A)dS spacetime as well as for its higher-dimensional generalizations with \cite{Gibbons:2004uw, Gibbons:2004js} or without \cite{MyersPerry:1986} the cosmological constant could be explicitly demonstrated \cite{Lunin:2017}. It is important to mention that such a separability occurs in special canonical coordinates. The very existence of these coordinates is intrinsically connected with a principal tensor, a special non-degenerate closed conformal Killing--Yano 2-form \cite{FrolovKrtousKubiznak:2017rewiew}. Such a tensor has been known to imply separability of the Hamilton--Jacobi, Klein--Gordon, and Dirac equations in higher-dimensional rotating black hole spacetimes {\cite{Frolov:2006pe,Sergyeyev:2007gf,OotaYasui:2008,Cariglia:2011qb,Cariglia:2011yt,CarigliaEtal:2012b}. There are some partial results also for other fields \cite{Kubiznak:2010ig,Oota:2008uj}, however, the link to the Maxwell equations remained hidden, cf.\ also \cite{FrolovKrtousKubiznak:2017rewiew}.}

In the present paper we uncover this connection and demonstrate that the separability ansatz proposed by Lunin for the potential of the electromagnetic field can be presented in a covariant way, in terms of the principal tensor. Such a method is therefore applicable to any spacetime admitting the principal tensor. In the present paper we focus on the case of four-dimensional metrics and demonstrate the separability of the Maxwell equations in the Kerr--NUT--(A)dS and the Pleba\'nski--Demia\'nski spacetimes. The higher-dimensional version of the covariant approach (based on the principal tensor) will be presented elsewhere \cite{KrtousEtal:2018}.

\section{From Pleba\'nski--Demia\'nski to off-shell canonical metric}

The Pleba\'nski--Demia\'nski solution \cite{PlebanskiDemianski:1976} describes the most general type D electro-vacuum
solution of Einstein--Maxwell equations with two commuting isometries. The ansatz takes the following form:
\be
\tilde{\ts{g}}=\Omega^{2}\ts{g}\hh \ts{\EMFbgr}=\grad\ts{\EMAbgr}\,,\n{eq1}
\ee
where

\begin{align}
\begin{split}
\ts{g}=&-\frac{\Delta_r}{\Sigma}(\grad\tau+y^2 \grad\psi)^2+\frac{\Delta_y}{\Sigma}(\grad\tau-r^2 \grad\psi)^2\\
       &+\frac{\Sigma}{\Delta_r}\grad r^2+\frac{\Sigma}{\Delta_y}\grad y^2\, ,\n{eq2}
\end{split}\\
\ts{\EMAbgr}=&-\frac{er}{\Sigma}\bigl(\grad \tau + y^2\,\grad\psi\bigr)
   -\frac{gy}{\Sigma}\bigl(\grad \tau - r^2\,\grad\psi\bigr)\;.\n{eq3}
\end{align}
Here,
\be\label{Sigmadef}
\Sigma=\sqrt{-g}=r^2+y^2\,,
\ee
and the conformal factor $\Omega$ reads
\begin{equation}\label{Omega_A}
\Omega^{\!-1}=1-yr\;.
\end{equation}
This ansatz obeys the Einstein--Maxwell equations with the electric and magnetic charges $e$ and $g$ and the cosmological constant $\Lambda$ provided the metric functions ${\Delta_y=\Delta_y(y)}$ and ${\Delta_r=\Delta_r(r)}$ take the following form:
\begin{equation}\label{functionsQP_A}
\begin{split}
\Delta_r&=k+e^2+g^2-2mr+\epsilon r^2-2nr^3-(k+\Lambda/3)r^4\;,\\
\Delta_y&=k+2ny-\epsilon y^2+2my^3-(k+e^2+g^2+\Lambda/3)y^4\,.
\end{split}
\end{equation}
Constants $k, m, \epsilon, n$ are free parameters that are related to mass, rotation, NUT parameter, and acceleration.  We refer to \cite{GriffithsPodolsky:2006b} for details and for a discussion and the interpretation of special cases of the Pleba\'nski--Demia\'nski metric.

The conformal metric (\ref{eq2}) is of its own interest. It gives rise to a vacuum solution of the Einstein equations provided one chooses the following metric functions $\Delta_r$ and $\Delta_y$:
\be\begin{split}\label{Delr}
\Delta_r&=(r^2+a^2)(1-\Lambda r^2/3)-2Mr\, ,\\
\Delta_y&=(a^2-y^2)(1+\Lambda y^2/3)+2Ny\, .
\end{split}
\ee
With these identifications, we recover the so-called Kerr--NUT--(A)dS metric \cite{Carter:1968cmp}, characterized by the mass $M$, rotation parameter $a$, and the NUT parameter $N$. $\Lambda$ as earlier is the cosmological constant.\footnote{Formally,
the Kerr--NUT--(A)dS geometry belongs to the Pleba\'nski--Demia\'nski class \eqref{eq1}. However, this is only obvious upon a proper redefinition of both coordinates and parameters, see \cite{GriffithsPodolsky:2006b}.}

In what follows we want to study the electromagnetic fields in the Pleba\'nski--Demia\'nski spacetime $\ts{\tilde g}$. Let us denote by $\ts{\EMA}$ the corresponding four-potential of the test Maxwell field.\footnote{Strictly speaking one must assume that $\ts{\EMAbgr}$ and $\ts{\EMA}$ are two independent vector fields. In the opposite case, if one considered $\ts{\EMA}$ as a perturbation of $\ts{\EMAbgr}$, the corresponding equations for such a perturbation would contain additions depending on the metric perturbations. In order to avoid this problem we always assume that $\ts{\EMA}$ is independent of $\ts{\EMAbgr}$ and obeys the standard (unmodified) Maxwell equations.}
Because of the conformal invariance of Maxwell's equations in four dimensions, the solutions $\tilde{\ts{\EMA}}$ and $\ts{\EMA}$ for the conformally related metrics $\tilde{\ts{g}}$ and $\ts{g}$ are identical. This suggests that for our calculations we might be able to use the Kerr--NUT--(A)dS metric $\ts{g}$, \eqref{eq2}, instead of the metric $\ts{\tilde g}$, \eqref{eq1}. (This is a very attractive idea as, as we shall see in the next section, the metric $\ts{g}$ admits a powerful extra symmetry that no longer exists for $\ts{\tilde g}$.) However, the two on-shell metrics are not simply conformally related, they have different metric functions $\Delta_r$ and $\Delta_y$, cf.~\eqref{functionsQP_A} and \eqref{Delr}. Nevertheless, as we shall see the separability property is directly linked to the principal tensor and prevails for any choice of metric functions $\Delta_y=\Delta_y(y)$ and $\Delta_r=\Delta_r(r)$ in \eqref{eq2}, and thus will also be valid for the special metric functions \eqref{functionsQP_A} in the metric~$\ts{\tilde g}$.

In other words, in order to demonstrate the separation of variables for the Maxwell equations in the Pleba\'nski--Demia\'nski metric, it is enough to show it for the metric~$\ts{g}$, \eqref{eq2}, with arbitrary metric functions $\Delta_y=\Delta_y(y)$ and $\Delta_r=\Delta_r(r)$. We shall call such a Kerr--NUT--(A)dS metric the {\em off-shell canonical metric}.

\section{Principal tensor}

The remarkable property of the off-shell canonical metric is that it admits a powerful symmetry encoded in the so called principal tensor. The principal tensor ${\ts{h}}$ is a {non-degenerate} closed conformal Killing--Yano 2-form. It is given by
\be
\ts{h}= y \grad y\wedge(\grad\tau-r^2 \grad\psi)
          -r \grad r\wedge (\grad\tau+y^2 \grad\psi)\,,
\ee
and obeys the equation
\be\label{PKYT}
\nabla_c h_{ab}=g_{ca}\xi_b-g_{cb}\xi_a\,,\quad \xi_a=\frac{1}{D-1}\nabla^b h_{ba}\, .
\ee
This tensor generates a number of explicit and hidden symmetries, and determines many remarkable properties of the geometry, see \cite{FrolovKrtousKubiznak:2017rewiew}.\footnote{Let us stress that the original Pleba\'nski--Demia\'nski metric \eqref{eq1} does not admit the principal tensor---one only has a much weaker conformal Killing--Yano tensor, given by $\tilde{\ts{h}}=\Omega^{3}\ts{h}$ \cite{Kubiznak:2007kh}.}

In particular, it is possible to show that $\ts{\xi}$ is a Killing vector, $\ts{\xi}=\cv{\tau}$. We call it a primary Killing vector. We also get a Killing--Yano tensor $\ts{f}=\ts{*h}$, and the associated conformal Killing tensor $\ts{Q}$ and Killing tensor $\ts{\kt}$, given by
\be
\ts{Q}=-\ts{h}\cdot \ts{h}\hh \ts{\kt}=-\ts{f}\cdot \ts{f}\,.
\ee
Here, the dot denotes a contraction of two tensors with respect to their nearby indices. That is, in components the previous relations take the following form:
\be
Q_{ab}=h_{ac} h_b^{\ c}\hh \kt_{ab}=f_{ac} f_b^{\ c}\,.
\ee
The vector $\ts{\zeta}=\ts{\kt}\cdot \ts{\xi}$ is a Killing vector, $\ts{\zeta}=\cv{\psi}$, and we call it a secondary Killing vector.

The Killing tensor $\ts{\kt}$ and the conformal Killing tensor $\ts{Q}$ obey the following properties:
\begin{gather}
\ts{\kt}-\ts{Q}=(r^2-y^2)\ts{g}\hh
\ts{\kt}\cdot\ts{Q}=r^2 y^2 \ts{g}\, ,\n{kkk}\\
\nabla_b \kt^b_{\ a}=-\frac12\nabla_a \kt^b{}_b\,,\n{dkk}
\end{gather}
see Eqs. (5.13) and (4.2) of \cite{FrolovKrtousKubiznak:2017rewiew}. Using relations \eqref{kkk}, one can show that
\be\label{pom22}
(\ts{g}+\mu^2 \ts{\kt})\cdot (\ts{g}-\mu^2 \ts{Q})=\Ag\ts{g}\, ,
\ee
where
\be
\Ag=q_r q_y\, ,\quad q_r=1+\mu^2 r^2\hh q_y=1-\mu^2 y^2\, .
\ee

Let us finally introduce the following four vectors:
\begin{align}
\ts{l}_{\pm}&=\cv{r}\pm {1\over \Delta_r}(r^2\cv{\tau}+\cv{\psi})\, ,\\
\ts{m}_{\pm}&=\cv{y}\pm {i\over \Delta_y}(-y^2\cv{\tau}+\cv{\psi})\, .
\end{align}

These vectors are null and have the following normalization:
\be
\ts{l}_+\!\cdot \ts{l}_-={2\Sigma\over \Delta_r}\, ,\quad
\ts{m}_+\!\cdot \ts{m}_-={2\Sigma\over \Delta_y}\,\, ,
\ee
while other scalar products vanish. (The normalization of the vectors $\ts{l}_{\pm}$ is chosen so that they are tangent to null geodesics in the affine parametrization.) They are the ``eigenvectors'' of the principal tensor
\be
\ts{h}\cdot\ts{l}_{\pm}=\pm r\, \ts{l}_{\pm}\hh
\ts{h}\cdot\ts{m}_{\pm}=\pm i y\, \ts{m}_{\pm}\, .
\ee
The corresponding eigenvalues $r$ and $y$ are two of the canonical coordinates of the metric (\ref{eq2}). The other two $(\tau, \psi)$ are the
Killing coordinates generated from $\ts{h}$ as described above. In other words, the canonical coordinates $(\tau, r, y, \psi)$ are uniquely determined by the principal tensor. It is in these coordinates the Maxwell equations will separate.

\section{Separability of Maxwell equations}

\subsection{Polarization tensor $\ts{B}$}

In order to construct a vector potential $\ts{\EMA}$ we shall use a special tensor $\ts{B}$, which we call a polarization tensor. We define it by the following relation:
\be\n{BHI}
(g_{ab}+i\mu h_{ab}) B^{bc}=\delta_a^c\,,
\ee
where $\mu$ is a (real) parameter related to the polarization of the electromagnetic wave.
{In the index-free notation the previous definition reads $(\ts{g}+i\mu \ts{h})\cdot \ts{B}=\ts{I}$.}

Using \eqref{pom22}, and denoting by
\be
\ts{\ktg}=\ts{g}+\mu^2 \ts{\kt}\, ,
\ee
we find that the polarization tensor $\ts{B}$ can be written as
\be\n{BOO}
\begin{split}
\ts{B}&=\frac{1}{\Ag}\, \ts{\ktg}\cdot (\ts{g}-i\mu \ts{h})\\
&=\frac{1}{\Ag}\bigl(\ts{g}-i\mu \ts{h}+\mu^2 \ts{\kt}-i\mu^3 \ts{\kt}\cdot \ts{h}\bigr)\,.
\end{split}
\ee
Let us emphasize that this is a quite non-trivial relation. The tensor $\ts{B}$ is defined as an inverse of a tensor which contains $\ts{h}$ linearly. {The relation (\ref{BOO}) shows that combination $\Ag\ts{B}$ can be written as a third order polynomial in~$\ts{h}$.}

\subsection{Field ansatz}

Let us consider the following ansatz for the electromagnetic field potential $\ts{\EMA}$:
\be \n{AAA}
\ts{\EMA}=\ts{B}\cdot\ts{\nabla}Z\,,
\ee
where $Z$ is a scalar function. We shall be looking for solutions of the Maxwell equations that admit the separation of variables in the following sense: the scalar function $Z$ is a product of four functions, each of which is a function of only one of the coordinates $(\tau,\psi,r,y)$
\be
Z=R(r) Y(y)E\hh E=e^{i\omega \tau} e^{i \tilde{m}\psi}\, .
\ee
The exponents which enter $E$ are eigen-functions of the {derivatives along} primary $\ts{\xi}=\cv{\tau}$ and secondary $\ts{\zeta}=\cv{\psi}$ Killing vectors\footnote{Let us mention that
for the Kerr metric the canonical coordinates $(\tau,\psi)$ differ from the standard time, $t$, and angle, $\phi$, coordinates by: $\tau=t-a \phi$ and $\psi=\phi/a$. One also has $y=a\cos\theta$. In these $(t,\phi)$ coordinates the function $E$ takes the form
\be
E=e^{i\omega t} e^{i m\phi}\, ,
\ee
where $m=a^{-1}\tilde{m} -a \omega$. The coordinate $\phi$ is periodic with a period $2\pi$. As a result the `quantum number' $m$ is `quantized' and takes integer values.}
{
\be\label{KVpush}
-i\lied_{\ts{\xi}}e^{i\omega \tau}=\omega\, e^{i\omega \tau}\hh
-i\lied_{\ts{\zeta}}e^{i\tilde{m} \psi}=\tilde{m}\, e^{i\tilde{m} \psi}\, .
\ee}

The ansatz (\ref{AAA}) for the potential is closely related to the one proposed by Lunin \cite{Lunin:2017}. In order to make the corresponding comparison, it is sufficient to write explicitly the components of the polarization tensor $\ts{B}$ in the frame $(\ts{l}_{\pm},\ts{m}_{\pm})$ in canonical coordinates $r$ and $y=a\cos\theta$. Upon substituting $\mu=(\mu_L a)^{-1}$ in $\ts{B}$, and up to a total constant normalization, the potential $\ts{\EMA}$ given by (\ref{AAA}) coincides with the ``magnetic mode'' (using the terminology of \cite{Lunin:2017}), where $\mu_L$ is the $\mu$-parameter used in Lunin's paper. The other, so called ``electric mode'' of Lunin's paper, coincides, up to a general constant factor,  with the expression (\ref{AAA}) provided one changes  $\ts{B}\to \ts{I}-\ts{B}$ and sets $\mu_L=-\mu$. However, a shift of $\ts{B}$ by a constant proportional to $\ts{I}$ is nothing but a pure gauge transformation that does not change the strength of the field $\ts{\EMF}$.\footnote{A natural generalization of the ansatz (\ref{BHI}) is a choice of the polarization tensor in the form
\be
\ts{B}={a \ts{I}+b\ts{h}\over c\ts{I}+d\ts{h}}\, .
\ee
However, the corresponding rational function of $\ts{h}$ can be written in the form $\ts{B}=C_0 \ts{I}+C_1/(\ts{I}+C_2\ts{h})$ with a proper choice of constants $C_i$. This implies that if one sets $C_2=i\mu$, the corresponding potential differs from (\ref{AAA}) by a pure gauge with coefficient $C_0$ and by a change of the normalization by factor $C_1$.}

After these general remarks let us return to our main problem. We shall proceed as follows.
We start from the field ansatz $\ts{\EMA}$ given by (\ref{AAA}), and impose the Lorenz gauge fixing condition. As a result we obtain a second order partial differential equation for the generating scalar function $Z$. We show that this equation can be solved by the separation of variables and derive the corresponding second-order ordinary differential equations (ODEs) for the mode functions $R(r)$ and $Y(y)$. Let us emphasize that this separation of variables is valid off-shell, that is for arbitrary functions $\Delta_r(r)$ and $\Delta_y(y)$ in \eqref{eq2}. Finally, by substituting the obtained solution $Z$ to the Maxwell field equations we demonstrate that they are identically satisfied.

\subsection{Lorenz gauge condition}

The Lorenz gauge condition reads
\be\n{LC}
\ts{\nabla}\cdot \ts{\EMA}=0\,.
\ee
By employing the ansatz \eqref{AAA} for the potential $\ts{\EMA}$, it can be written as
\be\label{coA}
\frac{1}{\Ag}\bigl[\ts{\nabla}\cdot(\ts{\ktg}\cdot\ts{\nabla}Z)-\ts{\nu}\cdot\ts{\ktg}\cdot\ts{\nabla}Z+i\mu \ts{W}\cdot \ts{\nabla}Z\bigr]=0\,,
\ee
or in components
\be\label{coA2}
\frac{1}{\Ag}\bigl[\nabla_a(\ktg^{ab}Z_{,b}) -\nu_a \ktg^{ab}Z_{,b}+i\mu W^a Z_{,a}\bigr]=0\,.
\ee
Here,
\begin{align}
\ts{\nu}&=\grad\ln \Ag=2\mu^2 \bigl({r\over q_r}\ts{d}r-{y\over q_y}\ts{d}y\bigr)\,,\\
\begin{split}
  \ts{W}&=\ts{\nu}\cdot\ts{\ktg}\cdot\ts{h}-\ts{\nabla}\cdot(\ts{\ktg}\cdot\ts{h})\\
        &=\bigl(1-{2\over q_r}-{2\over q_y}\bigr)(\ts{\xi} -\mu^2 \ts{\zeta})\,,
\end{split}
\end{align}
where $\ts{\xi}$ and $\ts{\zeta}$ are the primary and secondary Killing vectors. The last formula follows upon noting that
\be
\begin{aligned}
\ts{\nu}\cdot \ts{\ktg}&={2\mu^2\over \Sigma} \bigl({r\Delta_r q_y\over q_r}\ts{\pa}_r-{y\Delta_y q_r\over q_y}\ts{\pa}_y\bigr)\,,\\
\ts{\nu}\cdot \ts{\ktg}\cdot \ts{h}&={2\mu^2\over q_r q_y} \bigl[(\mu^4 r^2 y^2 +1)\ts{\zeta}\\
&\quad\quad -(2\mu^2 r^2 y^2 -r^2 +y^2)\ts{\xi} \bigr]\, ,\\
\ts{\nabla}\cdot (\ts{\ktg}\cdot \ts{h})&=3\ts{\xi}-\mu^2 \ts{\zeta}\,.
\end{aligned}
\ee
Using \eqref{KVpush}, the contribution of the last term in \eqref{coA2} is thus
\be\label{WWW}
i\mu W^a Z_{,a}=\mu\sigma\bigl(1-{2\over q_r}-{2\over q_y}\bigr)Z\,,
\ee
where
\be
\sigma=\mu^2 \tilde m-\omega\,.
\ee

Let us now turn to the first term. We shall denote  by prime a derivative with respect to $r$, and by dot a derivative with respect to $y$, we also define
\be
{\cal R}_1={R'\over R}\hhh {\cal R}_2={R''\over R}\hhh {\cal Y}_1={\dot{Y}\over Y}\hhh {\cal Y}_2={\ddot{Y}\over Y}\, .
\ee
Using these notations, we find
\be
\nabla_a(\ktg^{ab}Z_{,b})=
   \frac{\Ag}{\Sigma}\biggl(\frac{\mathcal{X}_0}{q_r}+\frac{\mathcal{U}_0}{q_y}\biggr)Z\, ,\n{ZZ.1}
\ee
where
\begin{align}
\mathcal{X}_0&=\Delta_r{\cal R}_2+{\Delta_r'} {\cal R}_1
   +\frac{(\omega q_r+\sigma)^2}{\mu^4 \Delta_r}+C{q_r}\, ,\n{ZZ.2}\\
\mathcal{U}_0&=\Delta_y{\cal Y}_2+{\dot{\Delta}_y} {\cal Y}_1
   -\frac{(\omega q_y+\sigma)^2}{\mu^4 \Delta_y}-C{q_y}\,.\n{ZZ.3}
\end{align}
The last terms in (\ref{ZZ.2}) and (\ref{ZZ.3}), proportional to an arbitrary constant $C$, reflects an ambiguity in the choice of ${\cal X}$ and ${\cal U}$ in the expression (\ref{ZZ.1}).

Adding the linear in derivatives term $-\nu_a O^{ab}Z_{,b}$ to
$\nabla_a(O^{ab}Z_{,b})$ results in the following changes in expressions for ${\cal X}_0$ and ${\cal U}_0$, see (\ref{ZZ.2}) and (\ref{ZZ.3})
\be
{\Delta_r'}\to {\Delta_r'}-\frac{2\mu^2 r}{q_r}\Delta_r\, ,\quad
{\dot{\Delta}_y}\to {\dot{\Delta}_y}+\frac{2\mu^2 y}{q_y}\Delta_y\, .
\ee
It is also possible to check that adding the term $i\mu W^c Z_{,c}$, \eqref{WWW}, to the obtained quantity results in the addition of the following terms to ${\cal X}_0$, ${\cal U}_0$, respectively:
\be
\frac{2-q_r}{\mu  q_r}\sigma\,,\quad
-\frac{2-q_y}{\mu  q_y}\sigma\,.
\ee

Thus we found that the Lorenz condition \eqref{LC} for the ansatz \eqref{AAA} can be written in the following form:
\be
\ts{\nabla}\cdot \ts{\EMA}=
\frac{Z}{\Sigma}\biggl(\frac{\mathcal{X}}{q_r}+\frac{\mathcal{U}}{q_y}\biggr)=0\,,
\ee
where
\begin{align}
\begin{split}
\mathcal{X}= \Delta_r{\cal R}_2&+\bigl({\Delta_r'}-\frac{2\mu^2 r}{q_r}\Delta_r\bigr){\cal R}_1 \\
       &+\frac{(\omega q_r+\sigma)^2}{\mu^4 \Delta_r}+\frac{2-q_r}{\mu  q_r}\sigma+C{q_r}\, ,
\end{split}\n{ZZ.4}\\
\begin{split}
\mathcal{U}= \Delta_y{\cal Y}_2&+\bigl({\dot{\Delta}_y}+ \frac{2\mu^2 y}{q_y}\Delta_y\bigr) {\cal Y}_1\\
       &-\frac{(\omega q_y+\sigma)^2}{\mu^4 \Delta_y}-\frac{2-q_y}{\mu  q_y}\sigma-C{q_y}\,.
\end{split}\n{ZZ.5}
\end{align}
Of course, since ${\cal X}={\cal X}(r)$ and ${\cal U}={\cal U}(y)$, the above requirement implies that we have to have ${\cal X}=0={\cal U}$\,. These equations can be written in the following explicit form:
\begin{gather}
\biggl(\frac{\Delta_r}{q_r}R'\biggr)'
  +\biggl(\frac{(\omega q_r+\sigma)^2}{\mu^4 \Delta_r q_r}+\frac{2-q_r}{\mu q_r^2}\sigma+C\biggr) R=0\, ,\n{RRYY}\\
\biggl(\frac{\Delta_y}{q_y}\dot{Y}\biggr)^{\cdot}
  -\biggl(\frac{(\omega q_y+\sigma)^2}{\mu^4 \Delta_y q_y}+\frac{2-q_y}{\mu q_y^2}\sigma+C\biggr) Y=0\, .\n{RRYY4}
\end{gather}

Let us notice that the parameter $\tilde{m}$ enters these equations only together with other parameters in a special combination, the quantity $\sigma$. Hence, the same separated equations are valid also in the Boyer--Lindquist type coordinates $(t,\phi)$ provided one sets
\be
\sigma=\mu^2 a(m+a\omega)-\omega=\mu^2 a^2 m-\omega(1-\mu^2 a^2)\, .
\ee

\subsection{Field equations}
Let us now turn to the Maxwell field equations:
\be\label{Maxwell}
\EMF_{ab}=2\EMA_{[b,a]}\hh J^a=\EMF^{ab}{}_{;b}\, .
\ee
In particular, we are interested in the source free fields, for which the current $\ts{J}$ vanishes, yielding
\be
(\sqrt{-g}\,\EMF^{ab})_{,b}=0\, .
\ee

Let us first discuss general properties of $\ts{J}$ for the field ansatz (\ref{AAA}). We split the coordinates into two groups
\be
y^\nu=(r,y)\, ,\quad \psi^{j}=(\tau,\psi)\,,
\ee
with indices taking mnemonic values ${\nu,\kappa,\ldots=r,y}$ and ${i,j,\ldots=\tau,\psi}$.
It is easy to see that for the canonical metric \eqref{eq2} terms with mixed indices vanish, $g_{\nu j}=g^{\nu j}=0$.  One has
\be\n{JJi}
J^\nu=(\sqrt{-g} F^{\nu\kappa})_{,\kappa}+\sqrt{-g}F^{\nu j}{}_{,j}\, .
\ee
Denote by $\epsilon^{\nu\kappa}$ a two-dimensional antisymmetric object,
\be
\epsilon^{\nu\kappa}=\epsilon^{[\nu\kappa]}\,\quad
\epsilon^{ry}=1\;.
\ee
and
\be
n_{j}=i(\omega \delta_{j}^{\tau}+\tilde{m}\delta_{j}^{\psi})\;.
\ee
Then one can rename components of the electromagnetic field as
\be
{\cal F}=\sqrt{-g}\,\EMF^{ry}\,,\quad
{\cal P}^\nu=\sqrt{-g}\, \EMF^{\nu j} n_j\, .
 \ee
In terms of these, conditions $J^\nu=0$ take the form
\be
\epsilon^{\nu\kappa} {\cal F}_{,\kappa}+{\cal P}^\nu=0\,.
\ee

Calculations give
\be
\begin{split}
{\cal F}=-\frac{\mu Z}{\Ag \Sigma}&\Bigl[
        \,q_r \Delta_r y(\tilde{m}{-}\omega y^2) {\cal R}_1
        +q_y \Delta_y r (\tilde{m}{+}\omega r^2) {\cal Y}_1\\
        &\,-\mu \Sigma\Delta_r \Delta_y {\cal R}_1 {\cal Y}_1\Bigr]\,,
\end{split}\raisetag{5.5ex}
\ee
\begin{align}
{\cal P}^r&=-Z\frac{\mu r \Delta_r (\tilde{m}+\omega r^2)}{\Sigma q_r}\, {\cal R}_2+\ldots\, ,\\
{\cal P}^y&=+Z\frac{\mu y \Delta_y (\tilde{m}-\omega y^2)}{\Sigma q_y}\, {\cal Y}_2+\ldots\, .
\end{align}
Here, the dots denote terms which contain less than two derivatives.
These relations imply that the components $J^\nu$ of the field equations do not contain derivatives of the mode functions $R$ and $Y$ higher than the second ones. Solving the equations $J^r=J^y=0$ with respect to $ {\cal R}_2$ and $ {\cal Y}_2$ one obtains expressions which are identical with those that follow from equations ${\cal X}={\cal U}=0$, \eqref{ZZ.4} and \eqref{ZZ.5}, provided one imposes
\be
C=0\,.
\ee
The current $J^a$ obeys the conservation law $J^a{}_{;a}=0$ which can be written in the form
\be\n{JJ}
(\sqrt{-g} J^\nu)_{,\nu}+\sqrt{-g}\,n_{j} J^{j}=0\, .
\ee
When $J^\nu=0$, one has $n_{j} J^{j}=0$. The direct calculations show that each of the components $J^{j}$ vanishes identically.

The obtained results mean that the Maxwell field equations for the vector potential ansatz (\ref{AAA}) are satisfied if the separated equations \eqref{RRYY} and \eqref{RRYY4} with $C=0$ hold true. Provided these equations, the Lorenz gauge condition is also automatically valid for this ansatz.

\section{Discussion}

In this paper we have demonstrated that for the vector potential \eqref{AAA}, the Maxwell equations {on the background of} the off-shell canonical metric \eqref{eq2} can be solved by the method of separation of variables. The ansatz \eqref{AAA} is a covariant generalization of the ansatz proposed by Lunin \cite{Lunin:2017}. As shown in this paper, it can be written in terms of the principal tensor $\ts{h}$. The remarkable property of the polarization tensor ${\ts{B}}$ is that although it is a rational function of $\ts{h}$, the combination ${\Ag\ts{B}}$ is a third order polynomial in $\ts{h}$.

We imposed the Lorenz gauge condition on the potential $\ts{\EMA}$ and showed that it is satisfied provided the functions $R(r)$ and $Y(y)$ which enter the mode function $Z$ obey homogeneous second order ODEs \eqref{RRYY} and \eqref{RRYY4}. These equations include one arbitrary constant, $C$. If this constant vanishes, $C=0$, the Maxwell equations are also satisfied.

In particular, the separation equations for the Pleba\'nski--Demia\'nski metric can be obtained by substituting the expressions \eqref{functionsQP_A} for $\Delta_r$ and $\Delta_y$ into \eqref{RRYY} and \eqref{RRYY4}. Of course, the same equations are also valid for the Kerr--NUT--(A)dS metric, characterized by functions {\eqref{Delr}}.

The results on the separability of Maxwell equations, established in this paper for the four-dimensional off-shell canonical metric by direct calculations, can be generalized to a general higher-dimensional case. The proof of this is quite involved and uses many remarkable properties of the principal tensor. This derivation is presented in \cite{KrtousEtal:2018}.

\vfill

\section*{Acknowledgements}
\label{sc:acknowledgements}
\addcontentsline{toc}{section}{Acknowledgements}

V.F.\ thanks the Natural Sciences and Engineering Research Council of Canada (NSERC) and the Killam Trust for their financial support. He also thanks the Yukawa Institute for Theoretical Physics at Kyoto University, where this work was completed during the workshop YITP-T-17-02 "Gravity and Cosmology 2018", for its hospitality.
P.K.\ was supported by Czech Science Foundation Grant \mbox{17-01625S}, and thanks University of Alberta for hospitality.
D.K.\ acknowledges the Perimeter Institute for Theoretical Physics and the NSERC for their support. Research at Perimeter Institute is supported by the Government of Canada through the Department of Innovation, Science and Economic Development Canada and by the Province of Ontario through the Ministry of Research, Innovation and Science.


%

\end{document}